\documentclass[11pt,a4paper]{article}
\pdfoutput=1
\usepackage{jheppub}
\usepackage{tikz}
\usepackage{subcaption}
\newcommand{\bra}{\langle}
\newcommand{\ket}{\rangle}

\newcommand{\cR}{\mathcal{R}}
\newcommand{\cD}{\mathcal{D}}

\newcommand{\bbZ}{{\mathbb Z}}
\newcommand{\Fc}{F^{\rm c}}
\newcommand{\Fo}{F^{\rm o}}
\newcommand{\gs}{g_{\rm s}}
\newcommand{\open}{{\rm o}}
\newcommand{\dg}{{\tilde{g}}}
\newcommand{\hpsi}{\widehat{\psi}}
\newcommand{\zs}{z_*}
\newcommand{\xis}[1]{\xi_*^{(#1)}}
\begin{document}

\title{Genus expansion of open free energy in 2d topological gravity}

\author[a]{Kazumi Okuyama}
\author[b]{and Kazuhiro Sakai}

\affiliation[a]{Department of Physics, Shinshu University,\\
3-1-1 Asahi, Matsumoto 390-8621, Japan}
\affiliation[b]{Institute of Physics, Meiji Gakuin University,\\
1518 Kamikurata-cho, Totsuka-ku, Yokohama 244-8539, Japan}

\emailAdd{kazumi@azusa.shinshu-u.ac.jp, kzhrsakai@gmail.com}

\abstract{
We study open topological gravity in two dimensions,
or, the intersection theory on the moduli space of
open Riemann surfaces
initiated by Pandharipande, Solomon and Tessler.
The open free energy,
the generating function for the open intersection numbers,
obeys the open KdV equations and Buryak's differential equation
and is related by a formal Fourier transformation
to the Baker-Akhiezer wave function of the KdV hierarchy.
Using these properties we study the genus expansion of
the free energy in detail.
We construct explicitly the genus zero part of the free energy.
We then formulate a method of computing higher genus
corrections by solving Buryak's equation
and obtain them up to high order.
This method is much more efficient than our previous approach
based on the saddle point calculation.
Along the way we show that
the higher genus corrections are polynomials in variables that are
expressed in terms of genus zero quantities only, generalizing
the constitutive relation of closed topological gravity.
}

\maketitle

%%%%%%%%%%%%%%%%%%%%%%%%%%%%%%%%%%%%%%%%%%%%%%%%%%%%%%%%%%%%%%%%%%%%%%%%
\section{Introduction}\label{sec:introduction}
%%%%%%%%%%%%%%%%%%%%%%%%%%%%%%%%%%%%%%%%%%%%%%%%%%%%%%%%%%%%%%%%%%%%%%%%

Two-dimensional gravity is one of the simplest models of quantum gravity
which has been intensively studied for quite some time.
In \cite{Gross:1989vs,Gross:1989aw,Douglas:1989ve,Brezin:1990rb}
it was found that two-dimensional gravity
is described by a certain double-scaled random matrix model
(see \cite{Ginsparg:1993is} for a review).
Mathematically, two-dimensional gravity
corresponds to an intersection theory on the moduli space of
closed Riemann surfaces, as first conjectured by Witten
\cite{Witten:1990hr}
and proved by Kontsevich \cite{Kontsevich:1992ti}.
It is known that the free energy of two-dimensional gravity on closed
Riemann surfaces satisfies the KdV equations
\cite{Douglas:1989dd,Witten:1990hr,Kontsevich:1992ti}
and the Virasoro constraints \cite{Fukuma:1990jw,Dijkgraaf:1990rs}.
Recently, it is realized that this story holds for Jackiw-Teitelboim
(JT) gravity as well; Saad, Shenker and Stanford \cite{Saad:2019lba}
showed that JT gravity is described by a doubled-scaled matrix model
and it corresponds to a particular background of Witten-Kontsevich
topological gravity
\cite{Mulase:2006baa,Dijkgraaf:2018vnm,Okuyama:2019xbv}.

Recently, Pandharipande, Solomon and Tessler
\cite{Pandharipande:2014qya} initiated the study 
of open topological gravity,
i.e.~the intersection theory on the moduli space of
Riemann surfaces with boundary. See also
\cite{Buryak:2014dta,Buryak:2014apa,Buryak:2015eza,tessler,
      Alexandrov:2014zva,Alexandrov:2014gfa} for related works.
It is conjectured in \cite{Pandharipande:2014qya} and proved
in \cite{Buryak:2015eza}
that the open free energy $\Fo(s)$,\footnote{
The variable $s$ is related to the 't Hooft parameter $\lambda$
in our previous papers
\cite{Okuyama:2019xbv,Okuyama:2020ncd} by
$\lambda=\sqrt{2}s$.} or the generating function of the 
open intersection numbers, satisfies the open version of the KdV
equations and the Virasoro constraints.
As explained in \cite{Dijkgraaf:2018vnm}, open topological gravity
is physically realized by adding vector degrees of freedom to
the matrix model of two-dimensional gravity.
After integrating out the vector degrees of freedom, this amounts to
the insertion of the determinant operator $\det(\xi-M)$
to the matrix integral,
where $\xi$ is a parameter and $M$ is the random matrix.
The expectation value of this determinant operator
\begin{equation}
\begin{aligned}
 \psi(\xi)=e^{-\frac{1}{2\gs}V(\xi)}\bra \det(\xi-M)\ket
\end{aligned} 
\label{eq:psi-det}
\end{equation}
corresponds to the wavefunction of
the FZZT brane \cite{Fateev:2000ik,Teschner:2000md}.
Here $V(\xi)$ is the matrix model potential and 
$\gs$ is the genus counting parameter (denoted as $u$ in
\cite{Pandharipande:2014qya,Buryak:2014dta,Buryak:2014apa,
Buryak:2015eza}).
$\psi(\xi)$ 
is also identified as the Baker-Akhiezer (BA) function
of the KdV hierarchy \cite{Maldacena:2004sn}.

It is known that the exponential of the open free energy
$e^{\Fo(s)}$ and the BA function $\psi(\xi)$ are 
related by the formal Fourier transformation
\cite{Buryak:2015eza,Dijkgraaf:2018vnm}\footnote{
The inverse transformation of \eqref{eq:Fopsirel}
is considered in \cite{Buryak:2015eza}. Our \eqref{eq:Fopsirel} is 
equivalent to eq.(4.76) in \cite{Dijkgraaf:2018vnm}.}
\begin{align}
\label{eq:Fopsirel}
e^{\Fo(s)}=\int_{-\infty}^\infty d\xi e^{\frac{s\xi}{\gs}}\psi(\xi).
\end{align}
One can compute the small $\gs$ expansion of $\Fo(s)$
\begin{equation}
\begin{aligned}
 \Fo(s)=\sum_{\dg=0}^\infty\gs^{\dg-1}\Fo_\dg(s)
\end{aligned} 
\label{eq:Fo-expand}
\end{equation}
from the 
result of the WKB expansion of the BA function 
by evaluating the integral \eqref{eq:Fopsirel} by the saddle point
method. For instance, the leading term of the small $\gs$ expansion of 
$\psi(\xi)\approx e^{-\frac{1}{2\gs}V_{\text{eff}}(\xi)}$
is given by the so-called effective potential $V_{\text{eff}}(\xi)$.
In our previous paper \cite{Okuyama:2020ncd}, we obtained
the explicit form of $V_{\text{eff}}(\xi)$
for arbitrary background couplings $\{t_n\}$.
Then the leading term $\Fo_0(s)$ in \eqref{eq:Fo-expand} 
is given by the Legendre transform of $V_{\text{eff}}(\xi)$.
One can in principle continue this saddle point computation for
the higher order corrections in $\gs$, but the computation becomes
very cumbersome as the order of $\gs$ increases. 

It turns out that the small $\gs$ expansion of 
$\Fo(s)$ can be computed systematically by
recursively solving Buryak's equation \cite{Buryak:2014apa},
which is understood as the Fourier transform of
the Schr\"odinger equation of the KdV hierarchy.
This is based on the fact that $\Fo_{\dg\ge 2}(s)$
in \eqref{eq:Fo-expand} is written as
a polynomial in variables
which are expressed in terms of genus zero quantities only.
This is similar to the situation
in original Witten-Kontsevich topological gravity,
where the genus-$g(\ge 2)$ closed free energy
$\Fc_g$ is written as a polynomial in a certain basis 
\cite{Itzykson:1992ya,Eguchi:1994cx}.

This paper is organized as follows. In section~\ref{sec:review}
we briefly review closed and open topological gravities.
We also explain how the open KdV equations
and Buryak's equation are derived from the KdV hierarchy.
In section~\ref{sec:gexp}
we study the genus expansion of the open free energy.
We first compute it from the genus expansion of the BA function
by the saddle point calculation.
We next derive an explicit expression of
the genus zero open free energy.
We then formulate a method of computing the genus expansion
by solving Buryak's equation.
We conclude in section~\ref{sec:conclusion}
with discussions on the future directions.
Some details of the calculations are relegated to
the appendices~\ref{sec:deriveoKdV} and~\ref{sec:Laginv}.

%%%%%%%%%%%%%%%%%%%%%%%%%%%%%%%%%%%%%%%%%%%%%%%%%%%%%%%%%%%%%%%%%%%%%%%%
\section{Brief review of topological gravity}\label{sec:review}
%%%%%%%%%%%%%%%%%%%%%%%%%%%%%%%%%%%%%%%%%%%%%%%%%%%%%%%%%%%%%%%%%%%%%%%%

In this section we will briefly review the basics
and known results about closed and open topological gravities.
We will also explain how the open KdV equations
and Buryak's equation, which will be the main tools
of our study of the open free energy, are derived from
the KdV hierarchy. 

%%%
\subsection{Witten-Kontsevich closed topological gravity}\label{sec:WK}
%%%

In Witten-Kontsevich topological gravity
\cite{Witten:1990hr,Kontsevich:1992ti}
(see also \cite{Dijkgraaf:2018vnm})
observables are made up of the intersection numbers
\begin{align}
\label{eq:intersec}
\langle\tau_{d_1}\cdots\tau_{d_n}\rangle_{g,n}
=\int_{\overline{\cal M}_{g,n}}
 \psi_1^{d_1}\cdots\psi_n^{d_n},\qquad
d_1,\ldots,d_n\in\bbZ_{\ge 0}.
\end{align}
They are defined on a closed Riemann surface $\Sigma$ of genus $g$
with $n$ marked points $p_1,\ldots,p_n$.
We let ${\cal M}_{g,n}$ denote the moduli space of $\Sigma$
and $\overline{\cal M}_{g,n}$
the Deligne-Mumford compactification of ${\cal M}_{g,n}$.
Here $\tau_{d_i}=\psi_i^{d_i}$ and
$\psi_i$ is the first Chern class of the complex line bundle
whose fiber is the cotangent space to $p_i$.
The intersection numbers \eqref{eq:intersec} obey the selection rule
\begin{align}
\langle\tau_{d_1}\cdots\tau_{d_n}\rangle_{g,n}=0
\quad \mbox{unless}\quad d_1+\cdots+d_n=3g-3+n.
\label{eq:selection}
\end{align}
The generating function for the above intersection numbers
is defined as
\begin{align}
\label{eq:genF}
\begin{aligned}
\Fc(\{t_k\},\gs)&:=\sum_{g=0}^\infty \gs^{2g-2}\Fc_g(\{t_k\}),\qquad
\Fc_g(\{t_k\})
 :=\left\langle e^{\sum_{d=0}^\infty t_d\tau_d}\right\rangle_g.
\end{aligned}
\end{align}
We will call $\Fc$ the closed free energy.

It was conjectured \cite{Witten:1990hr}
and proved \cite{Kontsevich:1992ti}
that $e^{\Fc}$ is a tau function for the KdV hierarchy.
This means that
\begin{align}
\label{eq:defu}
u:=\gs^2\partial_0^2\Fc
\end{align}
satisfies the KdV equations
\begin{align}
\label{eq:KdVeqs}
\partial_k u = \partial_0\cR_{k+1},
\end{align}
where $\cR_k$ are the Gelfand-Dikii differential polynomials of $u$
\begin{align}
\label{eq:cRforms}
\cR_0=1,\quad
\cR_1=u,\quad
\cR_2=\frac{u^2}{2}+\frac{\gs^2\partial_0^2 u}{12},\quad
\cdots.
\end{align}
Here we have introduced the notation
\begin{align}
\label{eq:dkdef}
\partial_k :=\frac{\partial}{\partial t_k}.
\end{align}
For $k=1$, \eqref{eq:KdVeqs} gives the traditional KdV equation
\begin{align}
\label{eq:ordKdV}
\partial_1 u
 =u\partial_0 u+\frac{\gs^2}{12}\partial_0^3 u.
\end{align}
Integrating \eqref{eq:KdVeqs} once in $t_0$ we have
\begin{align}
\label{eq:FRrel}
\gs^2\partial_k\partial_0 \Fc = \cR_{k+1}.
\end{align}

It is well known (see e.g.~\cite{DiFrancesco:1993cyw})
that the KdV equations \eqref{eq:KdVeqs}
are obtained as the compatibility condition of
the Schr\"odinger equation
\begin{align}
\label{eq:Schr}
Q\psi&=\xi\psi
\end{align}
and the KdV flow equations
\begin{align}
\label{eq:KdVflow}
\partial_k\psi = M_k \psi,
\end{align}
where
\begin{align}
\begin{aligned}
Q&:=\frac{\gs^2}{2}\partial_0^2+u,\qquad
M_k:=\frac{(2Q)_+^{k+1/2}}{(2k+1)!!\gs}.
\end{aligned}
\end{align}
Here we have decomposed
$(2Q)^{k+1/2}=(2Q)_+^{k+1/2}+(2Q)_-^{k+1/2}$
and the subscript $+$ means that
$(2Q)_+^{k+1/2}$ contains only non-negative powers of $\partial_0$.
Indeed, \eqref{eq:KdVeqs} is recovered by using the relation
\begin{align}
\label{eq:MQcom}
[M_k,Q]&=\partial_0\cR_{k+1}.
\end{align}
The wave function $\psi$ that satisfies
\eqref{eq:Schr} and \eqref{eq:KdVflow}
is known as the Baker-Akhiezer function.

Another important constraint that the closed free energy $\Fc$
obeys is the string equation \cite{Douglas:1989dd}.
For Witten-Kontsevich gravity it is written as
\begin{align}
u-\sum_{k=0}^\infty t_k\cR_k=0.
\end{align}
The genus zero part of this string equation is written as
\begin{align}
u_0-I_0(u_0,\{t_k\})=0,
\end{align}
where $u_0$ is the genus-zero part of $u$
\begin{align}
\label{eq:defu0}
u_0&:=\partial_0^2 \Fc_0
\end{align}
and we have introduced the Itzykson-Zuber variables
\cite{Itzykson:1992ya}
\begin{align}
\label{eq:defIn}
I_n(v,\{t_k\})=\sum_{\ell=0}^\infty t_{n+\ell}\frac{v^\ell}{\ell!}
\quad (n\ge 0).
\end{align}
Throughout this paper $I_n$ without specifying its arguments
should always be understood as
\begin{align}
I_n=I_n(u_0,\{t_k\}).
\end{align}
It is also convenient to introduce the variable
\begin{align}
\label{eq:tdef}
t:=(\partial_0 u_0)^{-1}=1-I_1.
\end{align}
It was conjectured \cite{Itzykson:1992ya}
and proved \cite{Eguchi:1994cx,Zhang:2019hly} that
$F_g(\{t_k\})\ (g\ge 2)$ are polynomials
in $I_{n\ge 2}$ and $t^{-1}$.
This fact significantly helps us to compute
higher genus free energy $\Fc_g$.

An efficient way to compute $\Fc_g$ is as follows.
(See \cite{Okuyama:2019xbv} for a more detailed explanation.)
Let us expand $u$ as
\begin{align}
\label{eq:uexp}
u=\sum_{g=0}^\infty \gs^{2g}u_g,\qquad u_g=\partial_0^2\Fc_g.
\end{align}
$u_g$ can be computed by recursively
solving the KdV equation \eqref{eq:ordKdV}.
To do this, let us regard $t_{k\ge 2}$ as parameters
and consider
the change of variables from $(t_0,t_1)$ to $(u_0,t)$.
The differentials $\partial_{0,1}$
are then written in the new variables as\footnote{
This change of variables was originally introduced by Zograf
(see e.g.~\cite{Zograf:2008wbe}).}
\begin{align}
\label{eq:d01inyt}
\partial_0
 =\frac{1}{t}(\partial_{u_0}-I_2\partial_t),
\qquad
\partial_1=u_0\partial_0-\partial_t.
\end{align}
By expanding both sides of the equation in $\gs$
\eqref{eq:ordKdV} is written as the recursion relation
\begin{align}
\label{eq:urecrel}
\begin{aligned}
-\frac{1}{t}\partial_t(tu_g)
&=\sum_{h=1}^{g-1}u_{g-h}\partial_0 u_h
 +\frac{1}{12}\partial_0^3u_{g-1}
\quad (g\ge 1).
\end{aligned}
\end{align}
This is easily solved with the help of \eqref{eq:d01inyt}.
First few of $u_g$ are
\begin{align}
\begin{aligned}
u_1&=\frac{I_2^2}{12t^4}+\frac{I_3}{24t^3},\\
u_2&=\frac{49I_2^5}{288t^9}+\frac{11I_2^3I_3}{36t^8}
 +\frac{84I_2^2I_4+109I_2I_3^2}{1152t^7}
 +\frac{32I_2I_5+51I_3I_4}{2880t^6}
 +\frac{I_6}{1152t^5}.
\end{aligned}
\end{align}
As explained in \cite{Okuyama:2019xbv}
one can easily integrate $u_g$ twice in $t_0$
and obtain the well-known results \cite{Itzykson:1992ya}
\begin{align}
\label{eq:Fcg}
\begin{aligned}
\Fc_1
 &=-\frac{1}{24}\log t,\\
\Fc_2
 &=\frac{I_4}{1152t^3}+\frac{29I_2I_3}{5760t^4}
   +\frac{7I_2^3}{1440t^5}.
\end{aligned}
\end{align}
%

%%%
\subsection{Pandharipande-Solomon-Tessler open topological gravity}
%%%

Pandharipande, Solomon and Tessler proposed
an open analog of Witten-Kontsevich
topological gravity \cite{Pandharipande:2014qya}.
They introduced the open intersection numbers
\begin{align}
\label{eq:ointersec}
\langle\tau_{d_1}\cdots\tau_{d_n}\sigma^k\rangle_{\dg,n}^\open
=2^{-\frac{\dg+k-1}{2}}
 \int_{\overline{\cal M}_{\dg,k,n}}
e(E,s),\qquad
d_1,\ldots,d_n\in\bbZ_{\ge 0}.
\end{align}
The new insertion $\sigma$ corresponds to the addition
of a boundary marking and the power $k$ of $\sigma$ specifies
the number of boundary markings.
$e(E,s)$ is the relative Euler class \cite{Pandharipande:2014qya},
which is thought of as an open analog of
the Euler class $e(E)=\psi_1^{d_1}\cdots\psi_n^{d_n}$
used in \eqref{eq:intersec}.
$\overline{\cal M}_{\dg,k,n}$ denotes a suitable compactification 
of the moduli space ${\cal M}_{\dg,k,n}$
of Riemann surfaces with boundary
of doubled genus $\dg$ with
graded spin structures,\footnote{
For the notion of graded spin structures, see e.g.~\cite{tessler}.}
$n$ interior and $k$ boundary marked points.
The open intersection numbers \eqref{eq:ointersec}
obey the selection rule
\begin{align}
\langle\tau_{d_1}\cdots\tau_{d_n}\sigma^k\rangle_{\dg,n}^\open=0
\quad \mbox{unless}\quad 2\sum_{j=1}^n d_j=3\dg-3+k+2n.
\label{eq:oselection}
\end{align}
The generating function for the open intersection numbers
is defined as
\begin{align}
\label{eq:Fogexp}
\begin{aligned}
\Fo(s,\{t_k\},\gs)
 &:=\sum_{\dg=0}^\infty \gs^{\dg-1}\Fo_\dg(s,\{t_k\}),
\qquad
\Fo_\dg(s,\{t_k\})
 :=\left\langle e^{s\sigma+\sum_{d=0}^\infty t_d\tau_d}
   \right\rangle_\dg^\open.
\end{aligned}
\end{align}
We will call $\Fo$ the open free energy.

It was conjectured \cite{Pandharipande:2014qya}
and then proved \cite{Buryak:2015eza}
that $\Fo$ satisfies the open KdV equations
\begin{align}
\label{eq:oKdV}
\begin{aligned}
\frac{2n+1}{2}\partial_n\Fo
 =\gs\partial_s\Fo\partial_{n-1}\Fo
 +\gs\partial_s\partial_{n-1}\Fo
 +\frac{\gs^2}{2}\partial_0\Fo\partial_0\partial_{n-1}\Fc
 -\frac{\gs^2}{4}\partial_0^2\partial_{n-1}\Fc
\quad (n\ge 1).
\end{aligned}
\end{align}
In fact it is known \cite{Buryak:2014apa} that
$\Fo$ is fully determined by the above system of equations
with the initial condition
\begin{align}
\label{eq:Foinit}
\Fo\Big|_{t_{k\ge 1}=0}
 =\frac{1}{\gs}\left(\frac{s^3}{6}+t_0 s\right),
\end{align}
given the closed free energy $\Fc$.
Buryak proved that $\Fo$ further satisfies
another differential equation \cite{Buryak:2014apa}
\begin{align}
\label{eq:Buryak}
\partial_s\Fo
 =\gs\left[\frac{1}{2}(\partial_0\Fo)^2+\frac{1}{2}\partial_0^2\Fo
  +\partial_0^2\Fc\right].
\end{align}
These equations play a crucial role
in the study of $\Fo$ in this paper.

In \cite{Buryak:2014dta} Buryak constructed
an explicit expression for $e^{\Fo}$ in terms of $\Fc$.
In this sense an explicit form of $\Fo$ is known.
For many purposes, however, it is still useful to
express $\Fo$ in the form of genus expansion \eqref{eq:Fogexp} and
construct an explicit, closed expression of $\Fo_\dg(s,\{t_k\})$
for fixed $\dg$.
This is our primary goal in this paper.

%%%
\subsection{Open free energy and Fourier transform of BA function}
\label{sec:Fourier}
%%%

It is known \cite{Buryak:2015eza,Dijkgraaf:2018vnm} that
the exponential of the open free energy $\exp F^{\rm o}$ is related
to the BA function $\psi$
by the formal Fourier transformation \eqref{eq:Fopsirel}.
Using this relation one can show \cite{Buryak:2015eza} that
the open KdV equations \eqref{eq:oKdV}
and Buryak's equation \eqref{eq:Buryak}
are in fact derived from the KdV flow equations \eqref{eq:KdVflow}
and the Schr\"odinger equation \eqref{eq:Schr} respectively.
In what follows we will present a derivation in a manner
slightly different from \cite{Buryak:2015eza}.

Let $\hpsi$ denote the formal Fourier transform of
the BA function $\psi$
\begin{align}
\label{eq:hpsidef}
\hpsi(s):=\int_{-\infty}^\infty d\xi e^{\frac{s\xi}{\gs}}\psi(\xi).
\end{align}
In terms of $\hpsi$
the Schr\"odinger equation \eqref{eq:Schr} is written as
\begin{align}
\label{eq:Schrhat}
Q\hpsi=\gs\partial_s\hpsi.
\end{align}
On the other hand, \eqref{eq:Fopsirel} is written as
\begin{align}
\label{eq:hpsiFo}
\hpsi&=e^{\Fo}.
\end{align}
Substituting \eqref{eq:hpsiFo} into \eqref{eq:Schrhat}
one obtains
\begin{align}
\label{eq:expBuryak}
\frac{\gs^2}{2}\partial_0^2 e^{\Fo}+u e^{\Fo}
=\gs\partial_s e^{\Fo}.
\end{align}
Rewriting $u$ by \eqref{eq:defu} we immediately see that
this is equivalent to \eqref{eq:Buryak}.
We have thus seen that
Buryak's equation \eqref{eq:Buryak} is nothing but
the formal Fourier transform of the Schr\"odinger equation
\eqref{eq:Schr}.

Similarly, let us consider the Fourier transform
of the KdV flow equations \eqref{eq:KdVflow}.
It is clear that the same equations hold for $\hpsi$ as well
\begin{align}
\label{eq:hpsiflow}
\begin{aligned}
\partial_n \hpsi
 &=M_n \hpsi\\
 &=\frac{(2Q)^{n+\frac{1}{2}}_+}{(2n+1)!!\gs}\hpsi.
\end{aligned}
\end{align}
It is shown \cite{Buryak:2014dta} that these equations are satisfied
by $\hpsi$ in \eqref{eq:hpsiFo} with $\Fo$ obeying
the open KdV equations \eqref{eq:KdVflow}.
Conversely, we can directly derive
the open KdV equations \eqref{eq:KdVflow} from \eqref{eq:hpsiflow}
using \eqref{eq:hpsiFo}.
Since the derivation is rather technical,
we relegate it to Appendix~\ref{sec:deriveoKdV}.
We stress that the open KdV equations are equivalent to
the KdV flow equations \eqref{eq:KdVflow}
under the identification \eqref{eq:hpsiFo}.

%%%%%%%%%%%%%%%%%%%%%%%%%%%%%%%%%%%%%%%%%%%%%%%%%%%%%%%%%%%%%%%%%%%%%%%%
\section{Genus expansion of open free energy}\label{sec:gexp}
%%%%%%%%%%%%%%%%%%%%%%%%%%%%%%%%%%%%%%%%%%%%%%%%%%%%%%%%%%%%%%%%%%%%%%%%

In this section we will study the genus expansion of
the open free energy. We will first compute it from
the genus expansion of the BA function
by the saddle point calculation.
We will next derive a fully explicit expression of
the genus zero open free energy.
Finally, we will formulate a method of computing the genus expansion
by solving Buryak's equation,
which turns out to be much more efficient than the saddle point
calculation.

%%%
\subsection{Genus expansion of BA function}
%%%

We saw in \cite{Okuyama:2020ncd} that the BA function $\psi$
admits the following expansion
\begin{align}
\psi=e^{A},\quad A=\sum_{\dg=0}^\infty\gs^{\dg-1}A_\dg.
\end{align}
First few of $A_\dg$ are\footnote{
In \cite{Okuyama:2020ncd} the constant part of $A_1$
is fixed so that it fits well with the convention of
closed topological gravity.
In this paper we will use this degree of freedom later
for compensating the difference of the normalizations of
$e^{\Fo}$ and $\psi$, so that we can avoid putting
an inessential normalization factor
in \eqref{eq:Fopsirel}.}
\begin{align}
\label{eq:An}
\begin{aligned}
A_0
 &=-\frac{t z^3}{3}
   +\sum_{n=1}^\infty\frac{I_{n+1}}{(2n+3)!!} z^{2n+3},\\
A_1&=-\frac{1}{2}\log z+\mbox{const.},\\
A_2&=-\frac{5}{24t z^3}-\frac{I_2}{24t^2 z},\\
A_3&=\frac{5}{16t^2 z^6}
     +\frac{1}{t^3}\left(\frac{I_2}{8 z^4}+\frac{I_3}{48 z^2}\right)
     +\frac{I_2^2}{24t^4 z^2},
\end{aligned}
\end{align}
where we have introduced\footnote{
$A_\dg$, $v_\dg$ and $z$ in this paper are related to those in
our previous paper \cite{Okuyama:2020ncd}
by $A_\dg^{\rm here}=\left(\sqrt{2}\right)^{1-\dg}A_\dg^{\rm there}$,
$v_\dg^{\rm here}=\left(\sqrt{2}\right)^{1-\dg}v_\dg^{\rm there}$,
$z^{\rm here}=\sqrt{2}z^{\rm there}$.}
\begin{align}
 z := \sqrt{2(\xi-u_0)}.
\end{align}
$A_\dg$ can be computed up to any order by solving
the recursion relation for $v_\dg:=\partial_0 A_\dg$
\begin{align}
\begin{aligned}
v_\dg
 &=-\frac{1}{2v_0}
    \left(\partial_0 v_{\dg-1}+\sum_{k=1}^{\dg-1}v_k v_{\dg-k}
         +\left\{\begin{array}{ll}
                 2u_{\frac{\dg}{2}}&\mbox{($\dg$ even)}\\
                 0&\mbox{($\dg$ odd)}
	         \end{array}\right.
   \right),\quad (n\ge 2),\\
v_0&= z,\quad
v_1=\frac{1}{2t z^2}.
\end{aligned}
\end{align}
In \cite{Okuyama:2020ncd} we performed this computation
with special values of $t_k$ corresponding to the case of JT gravity,
but as advertised in \cite{Okuyama:2020qpm} it can be generalized
without any effort to the case of general values of $t_k$,
as we have seen above.

%%%
\subsection{Saddle point calculation and polynomial structure of
            free energy}\label{sec:saddle}
%%%

The Fourier transformation \eqref{eq:Fopsirel} enables us
to calculate the genus expansion \eqref{eq:Fo-expand} of $\Fo$
from that of $A=\log\psi$ just obtained above.
\eqref{eq:Fopsirel} is written as
\begin{align}
\label{eq:FoinA}
\begin{aligned}
e^{\Fo}
 &=\int_{-\infty}^\infty d\xi e^{\frac{s\xi}{\gs}+A}\\
 &=\int_{-\infty}^\infty d\xi
   e^{[s\xi+A_0(\xi)]\gs^{-1}+A_1(\xi)+A_2(\xi)\gs+{\cal O}(\gs^2)}.
\end{aligned}
\end{align}
As in \cite{Okuyama:2019xbv,Okuyama:2020ncd}
one can calculate $\Fo_\dg$ by the saddle point method.

The saddle point $\xi_*$ is given by the condition
\begin{align}
\label{eq:saddlecond}
\partial_\xi\left[s\xi+A_0(\xi)\right]\Big|_{\xi=\xi_*}=0.
\end{align}
This is equivalent to
\begin{align}
\label{eq:sinzs}
\begin{aligned}
s&=-\partial_\xi A_0\Big|_{\xi=\xi_*}\\
 &=t\zs-\sum_{n=1}^\infty\frac{I_{n+1}}{(2n+1)!!}\zs^{2n+1},
\end{aligned}
\end{align}
where
\begin{align}
\label{eq:zsxisrel}
\zs:=\sqrt{2(\xi_*-u_0)}\quad\Leftrightarrow\quad
\xi_*=\frac{\zs^2}{2}+u_0.
\end{align}
By using the Lagrange inversion theorem this is inverted as
(see Appendix~\ref{sec:Laginv})
\begin{align}
\label{eq:zsins}
\zs
 &=\sum_{\substack{j_a\ge 0\\[.5ex]
                   \sum_a j_a=k\\[.5ex]
                   \sum_a aj_a=n}}
 \frac{(2n+k)!}{(2n+1)!}\frac{s^{2n+1}}{t^{2n+k+1}}
 \prod_{a=1}^\infty\frac{I_{a+1}^{j_a}}{j_a!(2a+1)!!^{j_a}}.
\end{align}

As in \cite{Okuyama:2019xbv} let us introduce a new variable $\phi$ as
\begin{align}
\xi=\xi_*+\sqrt{\gs}\phi.
\end{align}
The integral \eqref{eq:FoinA} is then written as
\begin{align}
\label{eq:saddleexp}
\begin{aligned}
e^{\Fo}
 &=
  e^{\left[s\xi_*+A_0(\xi_*)\right]\gs^{-1}+A_1(\xi_*)}
   \int_{-\infty}^\infty \sqrt{\gs}d\phi
  e^{\frac{1}{2}\partial_{\xi_*}^2A_0(\xi_*)\phi^2
     +{\cal O}(\gs^{1/2})}.
\end{aligned}
\end{align}
By expanding the integrand in $\gs$,
the integral in $\phi$ can be performed order by order
as a Gaussian integral.
In fact, we did essentially the same calculation
in \cite{Okuyama:2019xbv} up to the order of $\gs^1$.
We thus immediately obtain
\begin{align}
\label{eq:Fon}
\begin{aligned}
\Fo_0
 &=s\xi_*+A_0(\xi_*)
  =s\xi_*-\int_{u_0}^{\xi_*}d\xi_*' s(\xi_*')\\
 &=\int_0^s ds'\xi_\ast(s'),\\
\Fo_1
 &=A_1(\xi_*)
   +\frac{1}{2}\log\frac{2\pi\gs}{-\partial_{\xi_*}^2A_0(\xi_*)}\\
 &=\frac{1}{2}\log\frac{\partial_s\xi_*}{\zs}
  =\frac{1}{2}\log\partial_s\zs,\\[1ex]
\Fo_2
 &=A_2(\xi_*)
 +\frac{1}{2}\left(\frac{A_{0*}^{(3)}}{3!}\right)^2\langle\phi^6\rangle
 +\left(\frac{A_{0*}^{(4)}}{4!}
       +\frac{A_{0*}^{(3)}A_{1*}^{(1)}}{3!}\right)\langle\phi^4\rangle
 +\frac{\bigl(A_{1*}^{(1)}\bigr)^2+A_{1*}^{(2)}}{2}\langle\phi^2\rangle
\\
 &=
 -\frac{5}{24t\zs^3}-\frac{I_2}{24t^2\zs}
 +\frac{\xis{3}}{8\bigl(\xis{1}\bigr)^2}
 -\frac{\bigl(\xis{2}\bigr)^2}{6\bigl(\xis{1}\bigr)^3}
 -\frac{\xis{2}}{4\zs^2\xis{1}}
 +\frac{5\xis{1}}{8\zs^4},
\end{aligned}
\end{align}
where we have introduced the notation
\begin{align}
A_{\dg *}^{(n)}:=\partial_\xi^n A_\dg\Big|_{\xi=\xi_*},\qquad
\xis{n}:=\partial_s^n\xi_*.
\end{align}
In the last equality in \eqref{eq:Fon}
we have used
\begin{align}
\begin{aligned}
A_{0*}^{(n)}
 &=-\left(\frac{1}{\xis{1}}\frac{\partial}{\partial s}\right)^{n-1} s
\qquad(n\ge 1),\\
\langle\phi^{2m}\rangle
&=\frac{\int_{-\infty}^\infty d\phi e^{\frac{1}{2}A_{0*}^{(2)}\phi^2}
        \phi^{2m}}
       {\int_{-\infty}^\infty d\phi e^{\frac{1}{2}A_{0*}^{(2)}\phi^2}}
 =\frac{(2m-1)!!}{\bigl(-A_{0*}^{(2)}\bigr)^m}
 =(2m-1)!!\bigl(\xis{1}\bigr)^m\qquad(m\ge 1).
\end{aligned}
\end{align}
We have fixed the constant part of $A_1$ in \eqref{eq:An} in such a way
that the initial condition \eqref{eq:Foinit} is satisfied.
Using this method one can in principle calculate $\Fo_\dg$
up to any order. However, this calculation gets quickly involved as
$\dg$ increases. We will propose an alternative, much more efficient
method of computing $\Fo_\dg$ in the following subsections.

An advantage of the above calculation is that
we can prove the polynomial structure of
the higher genus free energies $\Fo_{\dg\ge 2}$.
The expansion \eqref{eq:saddleexp} implies that
$\Fo_{\dg\ge 2}$ are polynomials
in $A_{0*}^{(n\ge 3)}$, $A_{1*}^{(n\ge 1)}$, $A_{\dg\ge 2\,*}^{(n\ge 0)}$
and $\langle\phi^{2m}\rangle\ (m\ge 1)$.
On the other hand, by using the polynomial structure
of the closed free energy reviewed in section~\ref{sec:WK},
it is easy to show that
$A_{\dg\ge 2}$ are polynomials in $t^{-1}$, $I_{k\ge 2}$ and $z^{-1}$.
Combining these two lemmas we arrive at the conclusion that
$\Fo_{\dg\ge 2}$ are polynomials in
the variables $t^{-1}$, $I_{k\ge 2}$, $\zs^{-1}$,
$\bigl(\xis{1}\bigr)^{-1}$ and $\xis{n\ge 1}$.

It is well known that closed topological gravity
exhibits the constitutive relation \cite{Dijkgraaf:1990nc},
i.e.~higher genus quantities are expressed in terms of
genus zero quantities only.
In the case of Witten-Kontsevich gravity
$\Fc_1$ is given as in \eqref{eq:Fcg}
and $\Fc_{g\ge 2}$ are
expressed as polynomials in the variables
$t^{-1}$ and $I_{k\ge 2}$, as we saw in section~\ref{sec:WK}.
These variables are expressed explicitly in terms of
genus zero quantities $\partial_0^n u_0\ (n\ge 1)$ \cite{Zhou:2014spa}.
Since $\zs$ and $\xis{n}=\partial_s^{n+1}\Fo_0$
are also genus zero quantities,\footnote{
See \eqref{eq:d0Fo0}. It is also possible
to express $\xis{n}$ in terms of $t_0$-derivatives only.
This is done by repeatedly using \eqref{eq:d0xidszs}.}
the form of $\Fo_1$ in \eqref{eq:Fon} and
the above polynomial structure of $\Fo_{\dg\ge 2}$
ensure that Pandharipande-Solomon-Tessler
open topological gravity exhibits a generalized constitutive relation.

%%%
\subsection{Genus zero open free energy}\label{sec:Fo0}
%%%

In the last subsection we have obtained
an explicit expression of $\Fo_0$:
By plugging \eqref{eq:zsxisrel} into
the second line of \eqref{eq:Fon} we have
\begin{align}
\label{eq:Fo0inzszs}
\Fo_0=u_0s+\frac{1}{2}\int_0^s ds'\zs(s')^2
\end{align}
with $\zs(s)$ given in \eqref{eq:zsins}.
As we will see below,
we can write down a more direct expression for $\Fo_0$
by using the relations among $\Fo_0$, $\xi_*$ and $\zs$
which follow from
the system of equations \eqref{eq:Buryak} and \eqref{eq:oKdV}.

Buryak's equation \eqref{eq:Buryak} at the order of $\gs^{-1}$ reads
\begin{align}
\label{eq:Buryakm1}
\partial_s\Fo_0=\frac{1}{2}\left(\partial_0\Fo_0\right)^2+u_0.
\end{align}
Note also that the second line of \eqref{eq:Fon} gives
\begin{align}
\label{eq:dsFo0}
\partial_s\Fo_0=\xi_*.
\end{align}
Comparing these with \eqref{eq:zsxisrel} one finds\footnote{
Note that $\zs$ is the uniformization coordinate on the spectral curve.
In the context of minimal string theory, it is known that 
$\zs$ is given by the $t_0$-derivative of
the disk amplitude $\Fo_0$ \cite{Maldacena:2004sn}.}
\begin{align}
\label{eq:d0Fo0}
\partial_0\Fo_0=\zs.
\end{align}

On the other hand, the open KdV equation
\eqref{eq:oKdV} for $n=1$ at the order of $\gs^{-1}$ reads
\begin{align}
\label{eq:oKdVm1}
\frac{3}{2}\partial_1\Fo_0
 =\partial_s\Fo_0\partial_0\Fo_0
 +\frac{1}{2}\partial_0\Fo_0\partial_0^2\Fc_0.
\end{align}
By using \eqref{eq:d01inyt}, \eqref{eq:Buryakm1}, \eqref{eq:d0Fo0}
and \eqref{eq:defu0}
this becomes
\begin{align}
-\partial_t\Fo_0=\frac{\zs^3}{3}.
\end{align}
Applying $\partial_s$ to both sides of the equation
and using again \eqref{eq:Buryakm1} and \eqref{eq:d0Fo0}
one obtains
\begin{align}
\begin{aligned}
-\partial_t\left(\frac{\zs^2}{2}+u_0\right)
 &=\zs^2\partial_s\zs,
\end{aligned}
\end{align}
which gives
\begin{align}
\label{eq:dtdsrel}
\begin{aligned}
-\partial_t\zs
 &=\zs\partial_s\zs\\
 &=\partial_s\xi_*\\
 &=\partial_s^2\Fo_0.
\end{aligned}
\end{align}
Hence, differentiating \eqref{eq:zsins}
once in $t$ and then integrating it twice in $s$,
one obtains
\begin{align}
\label{eq:Fo0s}
\Fo_0
 &=u_0s
  +\sum_{\substack{j_a\ge 0\\[.5ex]
                   \sum_a j_a=k\\[.5ex]
                   \sum_a aj_a=n}}
 \frac{(2n+k+1)!}{(2n+3)!}\frac{s^{2n+3}}{t^{2n+k+2}}
 \prod_{a=1}^\infty\frac{I_{a+1}^{j_a}}{j_a!(2a+1)!!^{j_a}}.
\end{align}
The integration constants have been fixed accordingly
so that \eqref{eq:Fo0s} matches with \eqref{eq:Fo0inzszs}.
We verified by series expansion in $s$ that 
\eqref{eq:Fo0s} and \eqref{eq:Fo0inzszs} are indeed
in perfect agreement.
Note that when $t_{k\ge 1}=0$, we have
$t_0=u_0,\ t=1,\ I_{k\ge 2}=0$ and thus the above $\Fo_0$
becomes
\begin{align}
\Fo_0=t_0 s+\frac{s^3}{6}.
\end{align}
This is consistent with the initial condition \eqref{eq:Foinit}.

One can recast \eqref{eq:Fo0s} into another simple form.
Using
\begin{align}
\frac{1}{t^{2n+k+2}}
 =(1-I_1)^{-2n-k-2}
 =\sum_{j_0=0}^\infty
  \frac{(2n+k+j_0+1)!}{(2n+k+1)!}\frac{I_1^{j_0}}{j_0!}
\end{align}
and rewriting $a+1,j_{\ell-1},k+j_0$ as $a,j_\ell,k$ respectively
one obtains
\begin{align}
\label{eq:Fo0sbis}
\Fo_0
 &=u_0s
  +\sum_{\substack{j_a\ge 0\\[.5ex]
                   \sum_a j_a=k\\[.5ex]
                   \sum_a (a-1)j_a=n}}
 \frac{(2n+k+1)!}{(2n+3)!}s^{2n+3}
 \prod_{a=1}^\infty\frac{I_a^{j_a}}{j_a!(2a-1)!!^{j_a}}.
\end{align}
This expression can be directly compared with
the result of \cite{Pandharipande:2014qya}.
By observing
\begin{align}
u_0=t_0+t_0\cdot {\cal O}(t_*),\qquad
I_n=t_n+t_0\cdot {\cal O}(t_*),
\end{align}
and recalling the definition \eqref{eq:Fogexp}
with the selection rule \eqref{eq:oselection},
it is easy to see that \eqref{eq:Fo0sbis} indeed
reproduces the Theorem 1.4 of
\cite{Pandharipande:2014qya}
\begin{align}
\label{eq:PSTtheorem}
\left\langle \tau_{d_1}\cdots \tau_{d_\ell}
 \sigma^{\sum_{i=1}^\ell 2(d_i-1)+3}\right\rangle_{0,\ell}^\open
 =\frac{\bigl(\sum_{i=1}^\ell 2d_i-\ell+1\bigr)!}
       {\prod_{i=1}^\ell (2d_i-1)!!},\qquad
d_1,\ldots,d_\ell\ge 1.
\end{align}
Note that \eqref{eq:Fo0sbis} contains not only this formula
but also all the information
about open intersection numbers involving the operator $\tau_0$.

%%%
\subsection{Recursion relation}
%%%

By substituting the genus expansions \eqref{eq:Fo-expand}
and \eqref{eq:uexp}
Buryak's equation \eqref{eq:Buryak}
at the order of $\gs^{\dg-1}\ (\dg\ge 1)$ is written as
\begin{align}
\label{eq:recrelFon}
\cD\Fo_\dg
 =\frac{1}{2}\sum_{k=1}^{\dg-1}\partial_0\Fo_{\dg-k}\partial_0\Fo_k
 +\frac{1}{2}\partial_0^2\Fo_{\dg-1}
 +\left\{\begin{array}{ll}
         u_{\frac{\dg}{2}}&\mbox{($\dg$ even)}\\[1ex]
         0                &\mbox{($\dg$ odd)}
	 \end{array}\right.,
\end{align}
where we have introduced the differential operator
\begin{align}
\begin{aligned}
\cD&:=\partial_s-\zs\partial_0.
\end{aligned}
\end{align}
\eqref{eq:recrelFon} can be viewed as a recursion relation:
one can recursively compute $\Fo_\dg$
if one is able to perform the integration $\cD^{-1}$
on the l.h.s.~of \eqref{eq:recrelFon}.
This is indeed feasible, as we will see below.

To do this, let us first study the operator $\cD$,
which has in fact several interesting properties.
For instance, one can show that
\begin{align}
\label{eq:cDprop}
\begin{aligned}
\cD\zs&=\frac{1}{t},\\
\cD\xi_*&=0,\\
\cD\xis{n-1}
 =\cD\partial_s^{n-1}\xi_*
 &=\frac{1}{2}\sum_{k=1}^{n-1}
   \left(\begin{array}{@{}c@{}} n\\[1ex] k \end{array}\right)
   \partial_s^{n-k}\zs\partial_s^k\zs
\quad (n\ge 2).
\end{aligned}
\end{align}
The first line of \eqref{eq:cDprop} follows from
\begin{align}
\zs\partial_0\zs
=\partial_0(\xi_*-u_0)=\partial_0\xi_*-\frac{1}{t}
\end{align}
and
\begin{align}
\label{eq:d0xidszs}
\partial_0\xi_*=\partial_s\zs,
\end{align}
which follows from \eqref{eq:d0Fo0} by differentiating both sides
of the equation in $s$.
The second line of \eqref{eq:cDprop}
also follows from \eqref{eq:d0xidszs}.
The third line of \eqref{eq:cDprop} can easily be shown by induction.

It is also useful to note that
\begin{align}
\label{eq:xiszsrel}
\begin{aligned}
\xis{n}
  =\partial_s^n\xi_*
 &=\frac{1}{2}\sum_{k=0}^{n}
   \left(\begin{array}{@{}c@{}} n\\[1ex] k \end{array}\right)
   \partial_s^{n-k}\zs\partial_s^k\zs\quad (n\ge 1),
\end{aligned}
\end{align}
which immediately follows from \eqref{eq:zsxisrel}.
This relation is important
because it enables us to express
$\xis{n\ge 1}$ in terms of
\begin{align}
\zs^{(n)}:=\partial_s^n\zs\quad (n\ge 1),
\end{align}
and vice versa.
For instance, $\zs^{(n\ge 1)}$ with small $n$
are expressed in terms of $\xis{n\ge 1}$ as
\begin{align}
\label{eq:zinxi}
\begin{aligned}
\zs^{(1)}&=\frac{\xi_*^{(1)}}{\zs},\\
\zs^{(2)}
 &=\frac{\xi_*^{(2)}}{\zs}
  -\frac{\bigl(\xi_*^{(1)}\bigr)^2}{\zs^3},\\
\zs^{(3)}
 &=\frac{\xi_*^{(3)}}{\zs}
  -\frac{3\xi_*^{(1)}\xi_*^{(2)}}{\zs^3}
  +\frac{3\bigl(\xi_*^{(1)}\bigr)^3}{\zs^5}.
\end{aligned}
\end{align}
Moreover, comparing \eqref{eq:xiszsrel} with \eqref{eq:cDprop}
one finds that
\begin{align}
\label{eq:xis-cDxis}
\xis{n}=\cD\xis{n-1}+\zs\zs^{(n)}\quad (n\ge 1).
\end{align}
Therefore, by using \eqref{eq:zinxi} and \eqref{eq:xis-cDxis}
one can express $\cD\xis{n\ge 1}$ as polynomials
in $\xis{k\ge 1}$ and $\zs^{-1}$:
\begin{align}
\begin{aligned}
\cD\xi_*^{(1)}&=\frac{\bigl(\xi_*^{(1)}\bigr)^2}{\zs^2},\\
\cD\xi_*^{(2)}
 &=\frac{3\xi_*^{(1)}\xi_*^{(2)}}{\zs^2}
  -\frac{3\bigl(\xi_*^{(1)}\bigr)^3}{\zs^4},\\
\cD\xi_*^{(3)}
 &=\frac{4\xi_*^{(1)}\xi_*^{(3)}}{\zs^2}
  +\frac{3\bigl(\xi_*^{(2)}\bigr)^2}{\zs^2}
  -\frac{18\bigl(\xi_*^{(1)}\bigr)^2\xi_*^{(2)}}{\zs^4}
  +\frac{15\bigl(\xi_*^{(1)}\bigr)^4}{\zs^6}.
\end{aligned}
\end{align}

On the other hand, to evaluate the r.h.s.~of \eqref{eq:recrelFon}
it is convenient to use
\begin{align}
\label{eq:d0tzs}
\begin{aligned}
\partial_0\zs&=\frac{\zs^{(1)}}{\zs}-\frac{1}{t\zs},\\
\partial_0\xi_*^{(n)}&=\zs^{(n+1)}\quad (n\ge 0).
\end{aligned}
\end{align}
Again using \eqref{eq:zinxi} one can express
these quantities as polynomials in
$t^{-1}$, $\zs^{-1}$ and $\xis{n\ge 1}$.
Hence, by using the low genus results \eqref{eq:Fon}
and the polynomial structure of $\Fo_{\dg\ge 2}$ 
derived in section~\ref{sec:saddle},
it is easy to see that all quantities appearing
in \eqref{eq:recrelFon} are expressed as polynomials in
the variables $t^{-1}$, $I_{k\ge 2}$, $\zs^{-1}$,
$\bigl(\xis{1}\bigr)^{-1}$ and $\xis{n\ge 1}$.

%%%
\subsection{Higher genus open free energy}
%%%

We are now in a position to
solve the recursion relation \eqref{eq:recrelFon}
and compute the higher genus free energy $\Fo_\dg$.
To begin with, we verified that
$\Fo_\dg$ with $\dg=0,1,2$
given in \eqref{eq:Fon} indeed satisfy
the recursion relation \eqref{eq:recrelFon} for $\dg=1,2$.
This is easily done by using various identities derived
in the last two subsections.
Moreover, based on the polynomial structure discussed above,
one can perform the integration $\cD^{-1}$
completely and determine $\Fo_\dg$ unambiguously
for $\dg\ge 2$.
The algorithm to solve \eqref{eq:recrelFon}
and obtain $\Fo_{\dg}$ from
the data of $\{\Fo_{\dg'}\}_{\dg'<\dg}$ is as follows:
\renewcommand{\theenumi}{\roman{enumi}}
\renewcommand{\labelenumi}{(\theenumi)}
\begin{enumerate}
\item Compute the r.h.s.~of \eqref{eq:recrelFon} using \eqref{eq:d0tzs}
and express it as a polynomial in the variables 
$t^{-1}$, $I_{k\ge 2}$, $\zs^{-1}$, $\bigl(\xis{1}\bigr)^{-1}$
and $\xis{n\ge 1}$.

\item Let $t^{-m}f(I_k,\zs,\xi_*^{(n)})$
denote the highest order part in $t^{-1}$
of the obtained expression. This part can arise only from
\begin{align}
\cD\left(-\frac{f(I_k,\zs,\xi_*^{(n)})}{(m-2)t^{m-2}\zs I_2}\right).
\end{align}
Therefore subtract this from the obtained expression.

\item Repeat the procedure (ii) down to $m=3$.
Then all the terms of order $t^{-2}$ automatically disappear
and the remaining terms are of order $t^{-1}$ or $t^{0}$.
Note also that the expression does not contain any $I_k$.

\item In the result of (iii),
collect all the terms of order $t^{-1}$
and let $t^{-1}\partial_{\zs}g(\zs,\xi_*^{(n)})$
denote the sum of them. This part arises from
\begin{align}
\cD g(\zs,\xi_*^{(n)}).
\end{align}
Therefore subtract this from the result of (iii).
The remainder turns out to be independent of $t$.

\item
In the obtained expression,
let
\begin{align}
\frac{h\bigl(\xis{n\ge 2}\bigr)}{\zs^2\bigl(\xis{1}\bigr)^m}
\end{align}
denote the part which is of order $\zs^{-2}$
as well as of the lowest order in $\bigl(\xis{1}\bigr)^{-1}$.
This part arises from
\begin{align}
\cD\left(
\frac{h\bigl(\xis{n\ge 2}\bigr)}{(m+1)\bigl(\xis{1}\bigr)^{m+1}}\right).
\end{align}
Therefore subtract this from the obtained expression.

\item Repeat the procedure (v) until the resulting expression vanishes.

\item By summing up all the above obtained
primitive functions we obtain $\Fo_\dg$.

\end{enumerate}
Using this algorithm we computed $\Fo_\dg$ for $\dg\le 15$.\footnote{
The data of $\Fo_\dg$ are available upon request to the authors.}
We verified that $\Fo_2$ computed by this algorithm
reproduces the result \eqref{eq:Fon} of our saddle point calculation.
For $\dg=3$ we obtain
\begin{align}
\begin{aligned}
\Fo_3&=
\frac{I_2^2}{24t^4\zs^2}
+\frac{I_2}{8t^3\zs^4}
+\frac{I_3}{48t^3\zs^2}
+\frac{5}{16t^2\zs^6}
+\frac{I_2\xis{2}}{48t^2\zs^3\xis{1}}
-\frac{I_2\xis{1}}{12t^2\zs^5}
+\frac{5\xis{2}}{16t\zs^5\xis{1}}
-\frac{15\xis{1}}{8t\zs^7}\\
&\hspace{1em}
-\frac{35\xis{2}}{16\zs^6}
+\frac{35\bigl(\xis{1}\bigr)^2}{8\zs^8}
-\frac{\xis{4}}{16\zs^2\bigl(\xis{1}\bigr)^2}
+\frac{5\xis{3}}{16\zs^4\xis{1}}
-\frac{3\bigl(\xis{2}\bigr)^2}{16\zs^4\bigl(\xis{1}\bigr)^2}
-\frac{\bigl(\xis{2}\bigr)^3}{4\zs^2\bigl(\xis{1}\bigr)^4}\\
&\hspace{1em}
+\frac{7\xis{3}\xis{2}}{24\zs^2\bigl(\xis{1}\bigr)^3}
+\frac{\xis{5}}{48\bigl(\xis{1}\bigr)^3}
-\frac{\bigl(\xis{3}\bigr)^2}{8\bigl(\xis{1}\bigr)^4}
-\frac{\xis{4}\xis{2}}{6\bigl(\xis{1}\bigr)^4}
+\frac{3\xis{3}\bigl(\xis{2}\bigr)^2}{4\bigl(\xis{1}\bigr)^5}
-\frac{\bigl(\xis{2}\bigr)^4}{2\bigl(\xis{1}\bigr)^6}.
\end{aligned}
\end{align}
%

%%%%%%%%%%%%%%%%%%%%%%%%%%%%%%%%%%%%%%%%%%%%%%%%%%%%%%%%%%%%%%%%%%%%%%%%
\section{Conclusions and outlook}\label{sec:conclusion}
%%%%%%%%%%%%%%%%%%%%%%%%%%%%%%%%%%%%%%%%%%%%%%%%%%%%%%%%%%%%%%%%%%%%%%%%

In this paper we have studied the small $\gs$ expansion
\eqref{eq:Fo-expand} of
the open free energy $\Fo(s)$ of topological gravity.
We have obtained the explicit form \eqref{eq:Fo0s}
of the genus zero part $\Fo_0$ of the free energy.
We have then argued that the higher order corrections $\Fo_\dg$ can be
computed systematically by solving Buryak's equation recursively.
We have demonstrated this computation explicitly for the first few
orders. We have also elucidated the polynomial structure of
$\Fo_{\dg\ge 2}$. We emphasize that our result of $\Fo_\dg$ holds
for arbitrary value of the couplings $\{t_n\}$. 
We have shown that $\Fo_\dg$ is written as a combination of
genus-zero quantities only, which can be thought of as an open analog of 
the constitutive relation
for closed topological gravity \cite{Itzykson:1992ya,Eguchi:1994cx}.

The existence of the polynomial structure established in this paper
is important in several respects.
From a practical viewpoint, it is this structure that enables us to
determine the free energy merely by solving a single,
simple differential equation rather than
an infinite number of KdV flow equations or Virasoro constraints.
From a philosophical perspective,
our study would serve as a simple example of Gromov-Witten theories
in which the polynomial structure of the closed sector 
naturally extends to the open sector.
It could uncover the existence of similar polynomial structures
in a broader class of open Gromov-Witten theories.
Since our derivation of the polynomial structure is simply based on
the relation \eqref{eq:Fopsirel} or \eqref{eq:FoinA}
between the open free energy and the BA function,
it is likely to be generalized at least to some cases
such as the theory of open $r$-spin intersection numbers
\cite{Buryak:2018ypm}
for which a similar relation is known \cite{Bertola:2014yka}.
We stress that the polynomial structure is of the type
known as the constitutive relation, i.e.~all the generators
are genus zero quantities. All these may suggest the possibility 
of an open analog of Givental's formalism
\cite{Givental:2000cka,Givental:2001aja,Coates:2001ewh},
namely a general framework of
determining higher genus open Gromov-Witten invariants
from genus zero ones.

There are several interesting open questions.
In \cite{Alexandrov:2017ysm}
a refinement of the open intersection numbers is presented,
where how boundary markings are partitioned between
the boundaries is taken into account.
It is conjectured \cite{Alexandrov:2017ysm} that
the refined open free energy,
or the associated partition function $\tau_N^{\rm o,ext}$,
is written in terms of the Kontsevich-Penner matrix integral,
generalizing the result established in the $N=1$ unrefined case
\cite{Alexandrov:2014zva, Alexandrov:2014gfa}.
A similar matrix integral that is
deformed by an additional parameter $N$ is considered
for the Brezin-Gross-Witten model
\cite{Mironov:1994mv, Alexandrov:2016kjl}.
Its genus expansion is computed both
in the finite $N$ regime and
in the 't Hooft regime: $N\gg 1,\ \gs\ll 1$ with $\gs N$ fixed
\cite{Alexandrov:2016kjl, Okuyama:2020qpm}.
In particular, the genus expansion
in the latter regime can be computed
by simply solving the KdV equation \cite{Okuyama:2020qpm}.
It is interesting to see if the refined open free energy
can similarly be computed in the 't Hooft regime
by the techniques developed in this paper.

In general, the small $\gs$ expansion of 
$\Fo$ in \eqref{eq:Fo-expand} is an asymptotic series and we expect
that $\Fo$ receives non-perturbative corrections in $\gs$.
Such corrections are physically interpreted as the effect of
the so-called ZZ-branes \cite{Zamolodchikov:2001ah}.
It would be interesting to find the general structure of 
the effect of ZZ-branes for the arbitrary background $\{t_n\}$.
It is known that \cite{Seiberg:2003nm} some of
the background $\{t_n\}$ exhibits a non-perturbative instability
and it does not lead to a well-defined theory.
It would be interesting to find the map of the ``swampland''
in the space of all two-dimensional topological gravities $\{t_n\}$.
In particular, it is argued that the JT gravity matrix model
suffers from such a non-perturbative instability \cite{Saad:2019lba}.
It is important to see if JT gravity is non-perturbatively
well-defined or not.

\acknowledgments
This work was supported in part by JSPS KAKENHI Grant
Nos.~19K03845 and 19K03856,
and JSPS Japan-Russia Research Cooperative Program.

%%%%%%%%%%%%%%%%%%%%%%%%%%%%%%%%%%%%%%%%%%%%%%%%%%%%%%%%%%%%%%%%%%%%%%%%
%%%%%%%%%%%%%%%%%%%%%%%%%%%%%%%%%%%%%%%%%%%%%%%%%%%%%%%%%%%%%%%%%%%%%%%%
\appendix

%%%%%%%%%%%%%%%%%%%%%%%%%%%%%%%%%%%%%%%%%%%%%%%%%%%%%%%%%%%%%%%%%%%%%%%%
\section{Derivation of open KdV equations}\label{sec:deriveoKdV}
%%%%%%%%%%%%%%%%%%%%%%%%%%%%%%%%%%%%%%%%%%%%%%%%%%%%%%%%%%%%%%%%%%%%%%%%

In this section we will derive the open KdV equations \eqref{eq:oKdV}
from the KdV flow equations \eqref{eq:hpsiflow}.

Let $n$ be a positive integer.
Since
\begin{align}
Q^{n+\frac{1}{2}}
 =Q\cdot Q^{n-\frac{1}{2}}
 =Q\left(Q^{n-\frac{1}{2}}_+ + Q^{n-\frac{1}{2}}_-\right),
\end{align}
we have
\begin{align}
\begin{aligned}
Q^{n+\frac{1}{2}}_+
 =\left(Q\cdot Q^{n-\frac{1}{2}}_+\right)_+ 
  +\left(Q\cdot Q^{n-\frac{1}{2}}_-\right)_+
 =Q\cdot Q^{n-\frac{1}{2}}_+
  +\left(Q\cdot Q^{n-\frac{1}{2}}_-\right)_+.
\end{aligned}
\end{align}
Therefore \eqref{eq:hpsiflow} is rewritten as
\begin{align}
\label{eq:hpsiflow2}
\begin{aligned}
\frac{2n+1}{2}\partial_n\hpsi
 &=Q\frac{(2Q)^{n-\frac{1}{2}}_+}{(2n-1)!!\gs}\hpsi
  +\left(Q\frac{(2Q)^{n-\frac{1}{2}}_-}{(2n-1)!!\gs}\right)_+\hpsi.
\end{aligned}
\end{align}
Using \eqref{eq:hpsiflow} again one finds that
the first term on the r.h.s.~of \eqref{eq:hpsiflow2}
gives $Q\partial_{n-1}\hpsi$.
On the other hand,
it is known that (see e.g.~\cite{DiFrancesco:1993cyw})
$Q^{n-\frac{1}{2}}_-$ has the structure
\begin{align}
Q^{n-\frac{1}{2}}_-
 =\frac{(2n-1)!!}{2^{n+\frac{1}{2}}\gs}
  \{\cR_n,\partial_0^{-1}\}+{\cal O}(\partial_0^{-3}),
\end{align}
from which \eqref{eq:MQcom} follows. By using this,
the second term on the r.h.s.~of \eqref{eq:hpsiflow2} becomes
\begin{align}
\begin{aligned}
\left(Q\frac{(2Q)^{n-\frac{1}{2}}_-}{(2n-1)!!\gs}\right)_+\hpsi
 &=\frac{1}{4}
   \left(\partial_0^2\{\cR_n,\partial_0^{-1}\}\right)_+\hpsi\\[-2ex]
 &=\frac{1}{4}\left((\partial_0^2\cR_n)\partial_0^{-1}
     +(\partial_0\cR_n)+2\partial_0\cR_n\right)_+\hpsi\\
 &=\frac{1}{4}
   \left((\partial_0\cR_n)\hpsi+2\partial_0(\cR_n\hpsi)\right)\\
 &=\frac{3}{4}(\partial_0\cR_n)\hpsi+\frac{1}{2}\cR_n\partial_0\hpsi.
\end{aligned}
\end{align}
Hence \eqref{eq:hpsiflow2} becomes
\begin{align}
\label{eq:hpsiflow3}
\begin{aligned}
\frac{2n+1}{2}\partial_n\hpsi
 &=Q\partial_{n-1}\hpsi
  +\frac{3}{4}(\partial_0\cR_n)\hpsi+\frac{1}{2}\cR_n\partial_0\hpsi\\
 &=\left(\frac{\gs^2}{2}\partial_0^2+u\right)
   \partial_{n-1}\hpsi
  +\frac{3}{4}(\partial_{n-1}u)\hpsi
  +\frac{\gs^2}{2}(\partial_0\partial_{n-1}\Fc)\partial_0\hpsi\\
 &=\partial_{n-1}\left(\frac{\gs^2}{2}\partial_0^2+u\right)\hpsi
  -\frac{1}{4}(\partial_{n-1}u)\hpsi
  +\frac{\gs^2}{2}(\partial_0\partial_{n-1}\Fc)\partial_0\hpsi\\
 &=\partial_{n-1}Q\hpsi
  +\frac{\gs^2}{2}(\partial_0\partial_{n-1}\Fc)\partial_0\hpsi
  -\frac{1}{4}(\partial_{n-1}u)\hpsi.
\end{aligned}
\end{align}
In the second equality
we have used \eqref{eq:KdVeqs} and \eqref{eq:FRrel}.
Substituting \eqref{eq:Schrhat} we have
\begin{align}
\begin{aligned}
\frac{2n+1}{2}\partial_n\hpsi
&=\gs\partial_s\partial_{n-1}\hpsi
  +\frac{\gs^2}{2}(\partial_0\partial_{n-1}\Fc)\partial_0\hpsi
  -\frac{\gs^2}{4}(\partial_0^2\partial_{n-1}\Fc)\hpsi.
\end{aligned}
\end{align}
Under the identification \eqref{eq:hpsiFo}
one sees that this is equivalent to
the open KdV equations \eqref{eq:oKdV}.

%%%%%%%%%%%%%%%%%%%%%%%%%%%%%%%%%%%%%%%%%%%%%%%%%%%%%%%%%%%%%%%%%%%%%%%%
\section{\mathversion{bold}Derivation of $\zs(s)$}\label{sec:Laginv}
%%%%%%%%%%%%%%%%%%%%%%%%%%%%%%%%%%%%%%%%%%%%%%%%%%%%%%%%%%%%%%%%%%%%%%%%

In this section we will derive \eqref{eq:zsins} from \eqref{eq:sinzs}.

Suppose that
$w$ is expressed as a function of $z$ given by the formal power series
\begin{align}
w=f(z)=\sum_{n=1}^\infty f_n\frac{z^n}{n!}
\end{align}
with $f_1\ne 0$.
According to the Lagrange inversion theorem,
the inverse function is given by
\begin{align}
z=g(w)=\sum_{n=1}^\infty g_n\frac{w^n}{n!}
\end{align}
with
\begin{align}
\begin{aligned}
g_1&=\frac{1}{f_1},\qquad
g_n=\frac{1}{f_1^n}\sum_{k=1}^{n-1}(-1)^k
(n+k-1)!
\sum_{\{j_\ell\}}\prod_{\ell=1}^{n-k}
  \frac{1}{j_\ell}\left(\frac{f_{\ell+1}}{(\ell+1)!f_1}\right)^{j_\ell}
\quad (n\ge 2),
\end{aligned}
\end{align}
where the second sum is taken over all sequences
$j_1,j_2,\ldots,j_{n-k}$ of non-negative integers such that
\begin{align}
\label{eq:jcond}
\begin{aligned}
j_1+j_2+\cdots+j_{n-k}&=k,\\
j_1+2j_2+\cdots+(n-k)j_{n-k}&=n-1.
\end{aligned}
\end{align}

In the present case we have
\begin{align}
\label{eq:fkvalue}
\begin{aligned}
w&=s,\qquad z=\zs,\\
f_1&=t,\qquad
\frac{f_n}{n!}=
\left\{\begin{array}{cll}
-\dfrac{1}{n!!}I_{\frac{n+1}{2}}
&&n=3,5,7,\ldots,\\[2ex]
0&& n=2,4,6,\ldots.
\end{array}\right.
\end{aligned}
\end{align}
Since $f_{\ell+1}$ with odd $\ell$ are absent,
the conditions \eqref{eq:jcond} reduce to
\begin{align}
\begin{aligned}
&j_2+j_4+\cdots+j_{2\lfloor\frac{n-k}{2}\rfloor}=k,\\
&2j_2+4j_4+\cdots
 +2\lfloor\tfrac{n-k}{2}\rfloor j_{2\lfloor\frac{n-k}{2}\rfloor}=n-1.
\end{aligned}
\end{align}
It is clear that the second condition is satisfied only if $n$ is odd.
This means that all $g_n$ with even $n$ vanish.
For odd $n(\ge 3)$ we have
\begin{align}
\begin{aligned}
g_n
 &=\frac{1}{t^n}\sum_{k=1}^{n-1}(-1)^k
  (n+k-1)!\sum_{\{j_{2a}\}}
  \prod_{a=1}^{\lfloor\tfrac{n-k}{2}\rfloor}
  \frac{1}{j_{2a}!}\left(-\dfrac{I_{a+1}}{(2a+1)!!t}\right)^{j_{2a}}\\
 &=\sum_{k=1}^{n-1}
  \frac{(n+k-1)!}{t^{n+k}}\sum_{\{j_{2a}\}}
  \prod_{a=1}^{\lfloor\tfrac{n-k}{2}\rfloor}
  \frac{1}{j_{2a}!}\left(\dfrac{I_{a+1}}{(2a+1)!!}\right)^{j_{2a}}.
\end{aligned}
\end{align}
Therefore
\begin{align}
\begin{aligned}
\zs
 &=\frac{s}{t}+\sum_{m=1}^\infty
    \frac{s^{2m+1}}{(2m+1)!}g_{2m+1}\\
 &=\frac{s}{t}+\sum_{m=1}^\infty
    \frac{s^{2m+1}}{(2m+1)!}
  \sum_{k=1}^{2m}
  \frac{(2m+k)!}{t^{2m+k+1}}
  \sum_{\{j_{2a}\}}
  \prod_{a=1}^{\lfloor\tfrac{2m-k+1}{2}\rfloor}
  \frac{1}{j_{2a}!}\left(\dfrac{I_{a+1}}{(2a+1)!!}\right)^{j_{2a}}\\
 &=\frac{s}{t}
  +\sum_{\substack{j_{2a}\ge 0\\[.5ex]
                   \sum_a j_{2a}=k\ge 1\\[.5ex]
                   \sum_a aj_{2a}=m}}
 \frac{(2m+k)!}{(2m+1)!}
 \frac{s^{2m+1}}{t^{2m+k+1}}
 \prod_{a=1}^\infty
 \frac{1}{j_{2a}!}\left(\dfrac{I_{a+1}}{(2a+1)!!}\right)^{j_{2a}}\\
 &=\sum_{\substack{j_{2a}\ge 0\\[.5ex]
                   \sum_a j_{2a}=k\\[.5ex]
                   \sum_a aj_{2a}=m}}
 \frac{(2m+k)!}{(2m+1)!}
 \frac{s^{2m+1}}{t^{2m+k+1}}
 \prod_{a=1}^\infty
 \frac{1}{j_{2a}!}\left(\dfrac{I_{a+1}}{(2a+1)!!}\right)^{j_{2a}}.
\end{aligned}
\end{align}
By rewriting $j_{2a}$ as $j_a$ this gives \eqref{eq:zsins}.

%%%%%%%%%%%%%%%%%%%%%%%%%%%%%%%%%%%%%%%%%%%%%%%%%%%%%%%%%%%%%%%%%%%%%%%%
\bibliography{paper}
\bibliographystyle{utphys}

\end{document}